\documentclass [] {aa} 
\usepackage{epsfig, graphicx}

\begin{document}


\title{The metal-rich nature of stars with planets\thanks{Based on
    observations collected at the La Silla Observatory, ESO (Chile),
    with the {\footnotesize CORALIE} spectrograph at the 1.2-m
    Euler Swiss telescope, and with the {\footnotesize FEROS} spectrograph    
    at the 1.52-m ESO telescope (Observing run 66.C-0116\,B), and using the UES 
    spectrograph at the 4-m William Hershel Telescope (WHT), at La Palma 
    (Canary Islands).}}


\author{N.C.~Santos\inst{1} \and G.~Israelian\inst{2} \and M.~Mayor\inst{1}} 

\offprints{Nuno C. Santos, \email{Nuno.Santos@obs.unige.ch}}

\institute{Observatoire de Gen\`eve, 51 ch.  des
	   Maillettes, CH--1290 Sauverny, Switzerland \and 
	   Instituto de Astrofisica de Canarias, E--38205 La Laguna, Tenerife, Spain}

\date{Received / Accepted } 

\titlerunning{The metal-rich nature of stars with planets} 


\abstract{
With the goal of confirming the metallicity ``excess'' present in stars 
with planetary-mass companions, we present in this paper a high-precision 
spectroscopic study of a sample of dwarfs included in the CORALIE extrasolar 
planet survey. The targets were chosen according to the basic criteria that 
1) they formed part of a limited volume and 2) they did not present the 
signature of a planetary host companion. A few stars with planets were also 
observed and analysed; namely, \object{HD\,6434}, \object{HD\,13445} (Gl\,86), 
\object{HD\,16141}, \object{HD\,17051} ($\iota$ Hor), 
\object{HD\,19994}, \object{HD\,22049} ($\epsilon$ Eri), \object{HD\,28185}, 
\object{HD\,38529}, \object{HD\,52265}, \object{HD\,190228}, \object{HD\,210277} and \object{HD\,217107}. For some of these objects there had been no 
previous spectroscopic studies.\\
The spectroscopic analysis was done using the same technique as in previous 
work on the metallicity of stars with planets, thereby permitting a direct 
comparison of the results. The work described in this paper  thus represents 
the first uniform and unbiased comparison between stars with and without 
planetary-mass companions in a volume-limited sample. The results show that 1) 
stars with planets are significantly metal-rich, and 2) that the source 
of the metallicity is most probably ``primordial''. The results presented 
here may impose serious constraints on  planetary system formation and 
evolution models. 
}

\maketitle

\section{Introduction}

One of the most exciting and promising results that became evident after the 
discovery of the first extrasolar planets is that their host stars seem to 
be very metal-rich when compared with dwarf stars in the solar neighbourhood 
(Gonzalez \cite{Gon98a}; Santos et al. \cite{San00} -- hereafter Paper~I; 
Gonzalez et al. \cite{Gon01}). This result, representing the only link between the presence 
of planets and a stellar photospheric feature, has been given two main explanations. The first, 
is based on the classical view that giant planets are formed by runaway accretion of 
gas on to a ``planetesimal'' having up to 10 Earth masses.
In such a case, we can expect that the higher  the proportion is of dust to gas 
in the primordial cloud (i.e. metals), and consequently in the resulting proto-planetary disc, 
the more rapidly and easily may planetesimals, and subsequently 
the now observed giant planets be built. 

Opposing to this view, it had been proposed that 
the observed metallicity ``excess'' may be related to the ``pollution''
of the convective envelope of the star by the infall of planets and/or planetesimals 
(e.g., Gonzalez \cite{Gon98a}; Laughlin \& Adams \cite{Lau97}; Laughlin \cite{Lau00}; 
Gonzalez et al. \cite{Gon01}). This pollution can be the result of
the ``complete'' inward migration of a planet on to the star,
the transfer of material from the disc to the star as a result of the migration process 
(Goldreich \& Tremaine \cite{Gol80}; Lin et al. \cite{Lin96}), or to the break-up
and infall of a planet(s) on to the surface of
 the star due to gravitational interactions with other companions 
(Rasio \& Ford \cite{Ras96}). The former point might be particularly important for the short-period systems (Queloz et al. \cite{Que00}).

\begin{table*}[t]
\caption[]{Volume-limited sample of stars without detected giant planets.}
\begin{tabular}{lcccrccccc}
\hline
\noalign{\smallskip}
Star & $T_\mathrm{eff}$ & $\log{g}$     & $\xi_t$        & [Fe/H] & N(\ion{Fe}{i}) & N(\ion{Fe}{ii}) & $\sigma(\ion{Fe}{i})$ & $\sigma(\ion{Fe}{ii})$ & Mass  \\
     & (K)             & (cgs) & (km\,s$^{-1}$) &        &                &                 &                                          &                        & [$M_{\sun}$]\\\hline \\
\object{HD\,1581 }  &5940 &4.44 &1.13 &$-$0.15 &31 &7 &0.04 &0.05 &0.99$\pm$0.02\\
\object{HD\,4391 }  &5955 &4.85 &1.22 &0.01 &36 &5 &0.05 &0.09 &1.22$\pm$0.04\\
\object{HD\,5133 }  &5015 &4.82 &0.92 &$-$0.08 &36 &6 &0.05 &0.07 &0.81$\pm$0.04\\
\object{HD\,7570 }  &6135 &4.42 &1.46 &0.17 &35 &7 &0.04 &0.06 &1.19$\pm$0.02\\
\object{HD\,10360}  &5045 &4.77 &0.89 &$-$0.19 &36 &5 &0.04 &0.04 &0.70$\pm$0.03\\
\object{HD\,10647}  &6130 &4.45 &1.31 &$-$0.03 &34 &7 &0.03 &0.05 &1.14$\pm$0.03\\
\object{HD\,10700}  &5370 &4.70 &1.01 &$-$0.50 &38 &6 &0.04 &0.04 &0.68$\pm$0.02\\
\object{HD\,14412}  &5410 &4.70 &1.01 &$-$0.44 &35 &6 &0.04 &0.01 &0.78$\pm$0.05\\
\object{HD\,17925}  &5220 &4.60 &1.44 & 0.08 &35 &6 &0.07 &0.04 &0.92$\pm$0.06\\
\object{HD\,20010}  &6240 &4.27 &2.23 &$-$0.20 &33 &6 &0.05 &0.08 &1.33$\pm$0.01\\
\object{HD\,20766}  &5770 &4.68 &1.24 &$-$0.20 &35 &7 &0.04 &0.04 &0.97$\pm$0.04\\
\object{HD\,20794}  &5465 &4.62 &1.04 &$-$0.36 &39 &7 &0.05 &0.05 &0.74$\pm$0.02\\
\object{HD\,20807}  &5865 &4.59 &1.28 &$-$0.22 &37 &7 &0.04 &0.04 &0.95$\pm$0.02\\
\object{HD\,23249}  &5135 &4.00 &1.12 &0.17 &36 &7 &0.06 &0.08 &0.84$\pm$0.01\\
\object{HD\,23356}  &5035 &4.73 &0.96 &$-$0.05 &36 &6 &0.06 &0.07 &0.83$\pm$0.03\\
\object{HD\,23484}  &5230 &4.62 &1.13 &0.10 &37 &6 &0.05 &0.07 &0.92$\pm$0.06\\
\object{HD\,26965A} &5185 &4.73 &0.75 &$-$0.26 &37 &5 &0.05 &0.03 &0.71$\pm$0.02\\
\object{HD\,30495 } &5880 &4.67 &1.29 &0.03 &37 &7 &0.04 &0.03 &1.11$\pm$0.04\\
\object{HD\,36435 } &5510 &4.78 &1.15 &0.03 &37 &6 &0.06 &0.03 &1.04$\pm$0.05\\
\object{HD\,38858 } &5750 &4.56 &1.22 &$-$0.22 &36 &7 &0.04 &0.02 &0.91$\pm$0.02\\
\object{HD\,39091 } &5995 &4.48 &1.30 &0.09 &37 &7 &0.04 &0.04 &1.10$\pm$0.02\\
\object{HD\,40307 } &4925 &4.57 &0.79 &$-$0.25 &37 &4 &0.06 &0.10 &0.76$\pm$0.07\\
\object{HD\,43162 } &5630 &4.57 &1.36 &$-$0.02 &35 &7 &0.05 &0.04 &0.99$\pm$0.04\\
\object{HD\,43834 }  &5620 &4.56 &1.10 &0.12 &38 &7 &0.05 &0.05 &0.96$\pm$0.02\\
\object{HD\,50281A} &4790 &4.75 &0.85 &0.07 &31 &4 &0.05 &0.12 &0.79$\pm$0.01\\
\object{HD\,53705 } &5810 &4.40 &1.18 &$-$0.19 &36 &7 &0.03 &0.03 &0.92$\pm$0.01\\
\object{HD\,53706 } &5315 &4.50 &0.90 &$-$0.22 &36 &7 &0.05 &0.08 &0.88$\pm$0.05\\
\object{HD\,65907A} &5940 &4.56 &1.19 &$-$0.29 &39 &7 &0.04 &0.04 &0.94$\pm$0.01\\
\object{HD\,69830 } &5455 &4.56 &0.98 &0.00 &38 &7 &0.05 &0.03 &0.89$\pm$0.06\\
\object{HD\,72673 } &5290 &4.68 &0.81 &$-$0.33 &38 &6 &0.04 &0.04 &0.78$\pm$0.06\\
\object{HD\,74576 } &5080 &4.86 &1.20 &0.04 &36 &5 &0.06 &0.05 &0.91$\pm$0.04\\
\object{HD\,76151 } &5825 &4.62 &1.08 &0.15 &38 &7 &0.03 &0.04 &1.09$\pm$0.04\\
\object{HD\,84117 } &6140 &4.35 &1.38 &$-$0.04 &34 &7 &0.04 &0.06 &1.14$\pm$0.01\\
\object{HD\,189567} &5750 &4.57 &1.21 &$-$0.23 &37 &7 &0.04 &0.05 &0.89$\pm$0.01\\
\object{HD\,191408A}&5025 &4.62 &0.74 &$-$0.51 &37 &4 &0.06 &0.09 &0.61$\pm$0.04\\
\object{HD\,192310 }&5125 &4.63 &0.88 &0.05 &36 &6 &0.06 &0.08 &0.78$\pm$0.04\\
\object{HD\,196761 }&5460 &4.62 &1.00 &$-$0.27 &38 &7 &0.05 &0.05 &0.81$\pm$0.03\\
\object{HD\,207129 }&5910 &4.53 &1.21 &$-$0.01 &36 &7 &0.04 &0.03 &1.03$\pm$0.02\\
\object{HD\,209100 }&4700 &4.68 &0.60 &0.01 &34 &3 &0.07 &0.06 &0.64$\pm$0.09\\
\object{HD\,211415 }&5925 &4.65 &1.27 &$-$0.16 &35 &7 &0.03 &0.04 &0.99$\pm$0.02\\
\object{HD\,216803 }&4647 &4.88 &0.90 &0.07 &28 &3 &0.07 &0.08 &0.77$^{+0.01}$\hspace{-0.7truecm}$_{-0.18}$\\
\object{HD\,222237 }&4770 &4.79 &0.35 &$-$0.22 &37 &3 &0.08 &0.08 & -- \\
\object{HD\,222335 }&5310 &4.64 &0.97 &$-$0.10 &33 &5 &0.05 &0.04 &0.86$\pm$0.05\\
\\
\noalign{\smallskip}
\hline
\end{tabular}
\label{tab1}
\end{table*}
\begin{table*}
\caption[]{Determined atmospheric parameters and [Fe/H] for a set of stars with planets and brown dwarf companions (bd).}
\begin{tabular}{lccccccl}
\hline
\noalign{\smallskip}
Star & $T_\mathrm{eff}$ & $\log{g}$       & $\xi_t$        & [Fe/H] & Instrument & Mass         & Planet \\
     & (k)              &  (cgs) & (km\,s$^{-1}$) &        &            & [$M_{\sun}$] & discovery paper\\
\hline \\
\object{HD\,1237  } &5555$\pm$50 &4.65$\pm$0.15 &1.50$\pm$0.08 & 0.11$\pm$0.08 & CORALIE &1.01$\pm$0.04& Naef et al. (\cite{Nae01})\\
\object{HD\,6434  } &5790$\pm$40 &4.56$\pm$0.20 &1.40$\pm$0.10 &$-$0.55$\pm$0.07 &FEROS &0.79$\pm$0.01 & Queloz et al. (\cite{Que01})\\
\object{HD\,13445 } &5190$\pm$40 &4.71$\pm$0.10 &0.78$\pm$0.10 &$-$0.20$\pm$0.06 &CORALIE & &             \\
\object{HD\,13445 } &5220$\pm$40 &4.70$\pm$0.10 &0.85$\pm$0.07 &$-$0.19$\pm$0.06 &FEROS &  &              \\
\object{HD\,13445 }(avg) &5205        &4.70          &0.82          &$-$0.20 & --&0.75$\pm$0.04        & Queloz et al. (\cite{Que00})\\
\object{HD\,16141 } &5805$\pm$40 &4.28$\pm$0.10 &1.37$\pm$0.09 &0.15$\pm$0.05 &FEROS &1.06$\pm$0.01    & Marcy et al. (\cite{Mar00})    \\
\object{HD\,17051 } &6225$\pm$50 &4.65$\pm$0.15 &1.20$\pm$0.08 &0.25$\pm$0.06 &FEROS &1.26$\pm$0.02    & K\"urster et al. (\cite{Kur00})\\
\object{HD\,19994 } &6165$\pm$40 &4.13$\pm$0.20 &1.49$\pm$0.10 &0.23$\pm$0.06 &CORALIE &&                 \\
\object{HD\,19994 } &6250$\pm$40 &4.27$\pm$0.10 &1.56$\pm$0.08 &0.30$\pm$0.06 &FEROS & &                  \\
\object{HD\,19994 }(avg) &6210  &4.20          &1.52          &0.26 &-- &1.34$\pm$0.01           & Queloz et al. (\cite{Que01})\\
\object{HD\,22049 } &5135$\pm$40 &4.70$\pm$0.10 &1.14$\pm$0.07 &$-$0.07$\pm$0.06 &CORALIE &0.85$\pm$0.04& Hatzes et al. (\cite{Hat00})\\
\object{HD\,28185}  &5705$\pm$40 &4.59$\pm$0.10 &1.09$\pm$0.06 &0.24$\pm$0.05 & CORALIE &0.98$\pm$0.07 & Mayor et al. (\cite{May01b})\\
\object{HD\,38529 } &5675$\pm$40 &4.01$\pm$0.15 &1.39$\pm$0.09 &0.39$\pm$0.06 &FEROS &1.52$\pm$0.05    & Fisher et al. (\cite{Fis00})\\
\object{HD\,52265 } &6075$\pm$40 &4.21$\pm$0.10 &1.31$\pm$0.07 &0.22$\pm$0.07 &CORALIE &               &                 \\
\object{HD\,52265 } &6120$\pm$50 &4.37$\pm$0.20 &1.31$\pm$0.06 &0.25$\pm$0.06 &FEROS &                 &                   \\
\object{HD\,52265 }(avg)&6100        &4.29          &1.31          &0.24 &-- &1.18$\pm$0.01            & Naef et al. (\cite{Nae01})\\
                        &            &              &              &     &   &                         & Butler et al. (\cite{But00})\\
\object{HD\,75289 } &6135$\pm$40 &4.43$\pm$0.20 &1.50$\pm$0.07 &0.27$\pm$0.06 &CORALIE &1.20$\pm$0.02  & Udry et al. (\cite{Udr00})\\
\object{HD\,82943 } &6025$\pm$40 &4.54$\pm$0.10 &1.10$\pm$0.07 &0.33$\pm$0.06 &CORALIE &1.15$\pm$0.05  & Naef et al. (\cite{Nae01})\\
\object{HD\,83443 } &5500$\pm$60 &4.50$\pm$0.20 &1.12$\pm$0.09 &0.39$\pm$0.09 &CORALIE &0.90$\pm$0.05  & Mayor et al. (\cite{May01})\\
\object{HD\,108147} &6265$\pm$40 &4.59$\pm$0.15 &1.40$\pm$0.08 &0.20$\pm$0.06 &CORALIE &1.27$\pm$0.02  & Pepe et al. (in prep.)  \\
\object{HD\,121504} &6090$\pm$40 &4.73$\pm$0.10 &1.35$\pm$0.08 &0.17$\pm$0.06 &CORALIE &1.19$\pm$0.03  & Queloz et al. (\cite{Que01})\\
\object{HD\,162020}$^{bd}$ &4830$\pm$80 &4.76$\pm$0.25& 0.72$\pm$0.12  &0.01$\pm$0.11 &CORALIE &--     & Udry et al. (in prep.)      \\
\object{HD\,168746} &5610$\pm$30 &4.50$\pm$0.15 &1.02$\pm$0.08 &$-$0.06$\pm$0.05&CORALIE &0.88$\pm$0.01& Pepe et al. (in prep.) \\
\object{HD\,169830} &6300$\pm$30 &4.04$\pm$0.20 &1.37$\pm$0.07 &0.22$\pm$0.05 &CORALIE &1.42$\pm$0.01  & Naef et al. (\cite{Nae01})\\
\object{HD\,190228} &5360$\pm$40 &4.02$\pm$0.10 &1.12$\pm$0.08 &$-$0.24$\pm$0.06 &WHT &0.84$\pm$0.01   & Sivan et al. (\cite{Siv01})\\
\object{HD\,202206}$^{bd}$ &5765$\pm$40 &4.75$\pm$0.20 &0.99$\pm$0.09 &0.37$\pm$0.07 &CORALIE &1.02$\pm$0.02& Udry et al. (in prep.)   \\
\object{HD\,210277} &5575$\pm$30 &4.44$\pm$0.10 &1.12$\pm$0.08 &0.23$\pm$0.05 &FEROS &0.94$\pm$0.01    & Marcy et al. (\cite{Mar99})  \\
\object{HD\,217107} &5660$\pm$50 &4.43$\pm$0.10 &1.04$\pm$0.07 &0.40$\pm$0.06 &CORALIE &&                 \\
\object{HD\,217107} &5655$\pm$40 &4.42$\pm$0.05 &1.11$\pm$0.08 &0.38$\pm$0.05 &FEROS & &                  \\
\object{HD\,217107}(avg) &5660        &4.42          &1.01          &0.39          &-- &0.97$\pm$0.05  & Fischer et al. (\cite{Fis98})\\
\\
\noalign{\smallskip}
\hline
\end{tabular}
\label{tab2}
\end{table*}

The idea that planets and/or planetary material might be engulfed by a star is 
to some extent supported by the recent detection of $^6$Li in the planet host 
star \object{HD\,82943} (Israelian et al. \cite{Isr01}) -- interpreted as a signal of the
accretion of a planet during the history of the star -- and probably by the detection of a 
significant difference in iron abundances in the very similar pair of dwarfs 
\object{16 Cyg\,A} and \object{16 Cyg\,B}, which also have very different Li contents 
(Laws \& Gonzalez \cite{Law01}; Gonzalez \cite{Gon98a}).

Besides the [Fe/H] differences, there is currently some debate about possible
anomalies concerning other elements (Paper~I; Gonzalez et al. 
\cite{Gon01}; Smith et al. \cite{Smi01}). But the relatively low number of exoplanets known, 
and 
possible systematics with respect to the samples do not permit 
  firm 
conclusions to be reached on the subject.

The observation that stars with planets\footnote{Here we are referring 
to the  known extrasolar planetary host stars, having Jupiter-like planets in
relatively short period orbits when compared to the giant planets in our own Solar 
System.} are particularly metal-rich has so far been
overshadowed by the restriction that in order 
to compare the metallicities of stars hosting planets with those of stars 
``without'' planets, authors had access only 
 to published metallicity studies of 
volume-limited samples of dwarfs in the solar neighbourhood (mainly that of Favata et al. 
(\cite{Fav97}),  the alternative being to construct a sample using less precise 
metallicities using photometric indices). This is inconvenient for a
number of reasons. 
First and most obviously, one cannot be sure if the sample used for comparison
 is really 
free from giant planets. Furthermore, the metallicities for the Favata et al. sample were 
determined using a much shorter line list for iron and
different sources of atmospheric parameters (spectroscopic vs. colours) -- see Paper~I. 
This particular point may introduce systematic errors, and one might
 expect that 
the difference between the two samples was simply due to a bias related to the method used to compute 
the metallicities.

\begin{table*}
\caption[]{Stars with giant planets or brown dwarf companions (bd) studied by other authors.}
\begin{tabular}{lcccccll}
\hline
\noalign{\smallskip}
Star & $T_\mathrm{eff}$ & $\log{g}$    & $\xi_t$        & [Fe/H] & Mass        & Reference        & Planet\\
     & (k)            & (cgs) & (km\,s$^{-1}$) &        & [$M_{\sun}$]& for spectroscopy & discovery paper \\
\hline \\
\object{HD\,9826  } &6140 &4.12 &1.35 &0.12 & 1.28$\pm$0.02 & Gonzalez \& Laws (\cite{Gon00})    & Butler et al. (\cite{But97})\\
\object{HD\,10697 } &5605 &3.96 &0.95 &0.16 & 1.10$\pm$0.01 & Gonzalez et al. (\cite{Gon01})    & Vogt et al. (\cite{Vog00})\\
\object{HD\,12661}  &5714 &4.45 &0.99 &0.35 & 1.01$\pm$0.02 & Gonzalez et al. (\cite{Gon01})    & Fisher et al. (\cite{Fis00})\\
\object{HD\,37124}  &5532 &4.56 &0.85 &$-$0.41& -- & Gonzalez et al. (\cite{Gon01})             & Vogt et al. (\cite{Vog00})\\
\object{HD\,46375}  &5250 &4.44 &0.80 &0.21 & -- & Gonzalez et al. (\cite{Gon01})               & Marcy et al. (\cite{Mar00}) \\
\object{HD\,75732A}  &5250 &4.40 &0.80 &0.45 & 1.05$\pm$0.03 & Gonzalez \& Vanture (\cite{Gon98b}) & Butler et al. (\cite{But97})\\
\object{HD\,80606}  &5645 &4.50 &0.81 &0.43 & $\sim$1.1 & Naef et al. (\cite{Nae01b}) & Naef et al. (\cite{Nae01b})\\
\object{HD\,89744}  &6338 &4.17 &1.55 &0.30 & 1.55$\pm$0.03 & Gonzalez et al. (\cite{Gon01})    & Korzennik et al. (\cite{Kor00})\\
\object{HD\,92788}  &5775 &4.45 &1.00 &0.31 & 1.05$\pm$0.02& Gonzalez et al. (\cite{Gon01})     & Queloz et al. (\cite{Que01})\\
                    &     &     &     &     &              &                                    & Fisher et al. (\cite{Fis00})\\
\object{HD\,95128}  &5800 &4.25 &1.0 &0.01 & 1.03$\pm$0.03 & Gonzalez (\cite{Gon98a})           & Butler \& Marcy (\cite{But96})\\
\object{HD\,114762}$^{bd}$ &5950 &4.45 &1.0 &$-$0.60 & 0.82$\pm$0.03 & Gonzalez (\cite{Gon98a}) & Latham et al. (\cite{Lat89})\\
\object{HD\,117176} &5500 &3.90 &1.0 &$-$0.03 & 1.10$\pm$0.02 & Gonzalez (\cite{Gon98a})        & Marcy \& Butler \cite{Mar96})\\
\object{HD\,120136} &6420 &4.18 &1.25 &0.32 & 1.34$\pm$0.02 & Gonzalez \& Laws (\cite{Gon00})   & Butler et al. (\cite{But97})\\
\object{HD\,130322} &5410 &4.47 &0.95 &0.05 & $\sim$0.89 & Gonzalez et al. (\cite{Gon01})        & Udry et al. (\cite{Udr00})\\
\object{HD\,134987} &5715 &4.33 &1.00 &0.32 & 1.02$^{+0.07}$\hspace{-0.7truecm}$_{-0.03}$ & Gonzalez et al. (\cite{Gon01})& Vogt et al. (\cite{Vog00})\\
\object{HD\,143761} &5750 &4.10 &1.2 &$-$0.29 & 0.96$\pm$0.03 & Gonzalez (\cite{Gon98a})         & Noyes et al. (\cite{Noy97})\\
\object{HD\,145675} &5300 &4.27 &0.80 &0.50 & $\sim$1.05 & Gonzalez et al. (\cite{Gon99})        & Udry et al (\cite{Udr01})\\
\object{HD\,168443} &5555 &4.10 &0.90 &0.10 & 1.01$\pm$0.02 & Gonzalez et al. (\cite{Gon01})     & Marcy et al. (\cite{Mar99})\\
                    &     &     &     &     &               &                                    & Udry et al. (\cite{Udr01})\\
\object{HD\,177830} &4818 &3.32 &0.97 &0.36 & 1.03$\pm$0.11 & Gonzalez et al. (\cite{Gon01})     & Vogt et al. (\cite{Vog00})\\
\object{HD\,186427} &5685 &4.26 &0.80 &0.07 & 0.97$\pm$0.03 & Laws \& Gonzalez (\cite{Law01})    & Cochran et al. (\cite{Coc97})\\
\object{HD\,187123} &5830 &4.40 &1.00 &0.16 & 1.08$\pm$0.04 & Gonzalez et al. (\cite{Gon99})     & Butler et al (\cite{But98})\\
\object{HD\,192263} &4964 &4.49 &0.95 &$-$0.03 & $\sim$0.80 & Gonzalez et al. (\cite{Gon01})     & Santos et al. (\cite{San00b})\\
                    &     &     &     &        &            &                                    & Vogt et al. (\cite{Vog00})\\
\object{HD\,209458} &6063 &4.38 &1.02 &0.04 & 1.12$\pm$0.02 & Gonzalez et al. (\cite{Gon01})     & Mazeh et al. (\cite{Mah00})\\   
                    &     &     &     &     &               &                                    & Henry et al. (\cite{Hen00})\\   
\object{HD\,217014} &5795 &4.41 &1.05 &0.21 & 1.07$\pm$0.01 & Gonzalez et al. (\cite{Gon01})     & Mayor \& Queloz (\cite{May95})\\
\object{HD\,222582} &5735 &4.26 &0.95 &0.02 & 0.95$\pm$0.01 & Gonzalez et al. (\cite{Gon01})     & Vogt et al. (\cite{Vog00})\\
\noalign{\smallskip}
\hline
\end{tabular}
\label{tab3}
\end{table*}

With the goal of settling  the question about the high 
metallicity content of stars with planets, we present here a spectroscopic study 
of a volume-limited sample of 43 stars included in the {\footnotesize CORALIE} 
(Udry et al. \cite{Udr00}) planet search programme, and for which the radial velocities 
seem to be constant over a large time interval. The technique used, line lists 
and atmospheric models were those usually applied by most authors working on 
the metallicities of stars with planets (e.g. Paper~I; Gonzalez 
et al. \cite{Gon01}). We show that the currently known stars with 
giant planets are on average more metal-rich than ``field stars'', for which there is no radial-velocity 
signature of planets. Furthermore, the results are used to set strong constraints on 
the cause of the observed ``anomaly'', significantly
excluding the 
``pollution'' scenario. The impact on the planetary formation and evolutionary models
is discussed.

\section{The data}

\subsection{The samples}

As mentioned above, the choice of a ``comparison sample'' (as we shall call it from 
now on), free from giant planets and with the minimum bias possible, is of crucial
importance. In order to be sure that we are not including any systematics when 
selecting the sample, we have taken all stars within the CORALIE sample
(defined as volume-limited -- Udry et al. \cite{Udr00}) having right ascensions between 
20$^\mathrm{h}$ and 9$^\mathrm{h}$ (observable in mid November from La Silla). Since such 
a sample corresponds to more than half of the $\sim$ 1600 stars searched for planets, 
it is virtually impossible to make a precise spectroscopic analysis for all those stars
in a reasonable amount time. We thus decided as a first step to limit our observations
to all stars within 17\,pc of the Sun having $(B-V)<1.1$ (determined from Hipparcos data -- 
ESA \cite{ESA97}), and within 
20\,pc with $(B-V)<0.9$. This sample includes about 50 stars, most of which were observed. 
Some of the stars in this sample have known 
planetary systems,  namely \object{HD\,1237}, \object{HD\,13445}, \object{HD\,17051}, 
\object{HD\,22049}, \object{HD\,28185} and \object{HD\,217107} (\object{HD\,192263}, would also 
have  been included if 
we had not imposed any limit on $(B-V)$). All these objects were excluded from our 
final sample (43 objects) presented in Table~\ref{tab1}.

In what concerns the stars with planets, all known objects having precise 
spectroscopic iron abundance determinations using a similar technique were used. This 
excludes \object{Gl\,876} (Delfosse et al. \cite{Del98}; Marcy et al. \cite{Mar98}), 
\object{HD\,195019} (Fischer et al. \cite{Fis98}) and the recently announced 
planets around \object{HD\,160691}, \object{HD\,27442} and \object{HD\,179949} (Tinney 
et al. \cite{Tyn01}; Butler et al. \cite{But01}). Also, from the set of 9 new planet host 
stars recently announced by the Geneva group (Mayor et al. \cite{May01b})
only two had available data (\object{HD\,28185} and \object{HD\,80606}). The only exception was made 
respecting \object{BD$-$10\,3166}, which was included in the planet surveys for its high [Fe/H] 
(Gonzalez et al. \cite{Gon99}). All the other stars were part of search programmes  for planets in
volume-limited samples (Udry et al. \cite{Udr00}; Marcy et al. \cite{Mar00}), and there is 
no particular reason to take out any of the targets (possible sources of bias are discussed below).

\begin{figure*}[t]
\psfig{width=\hsize,file=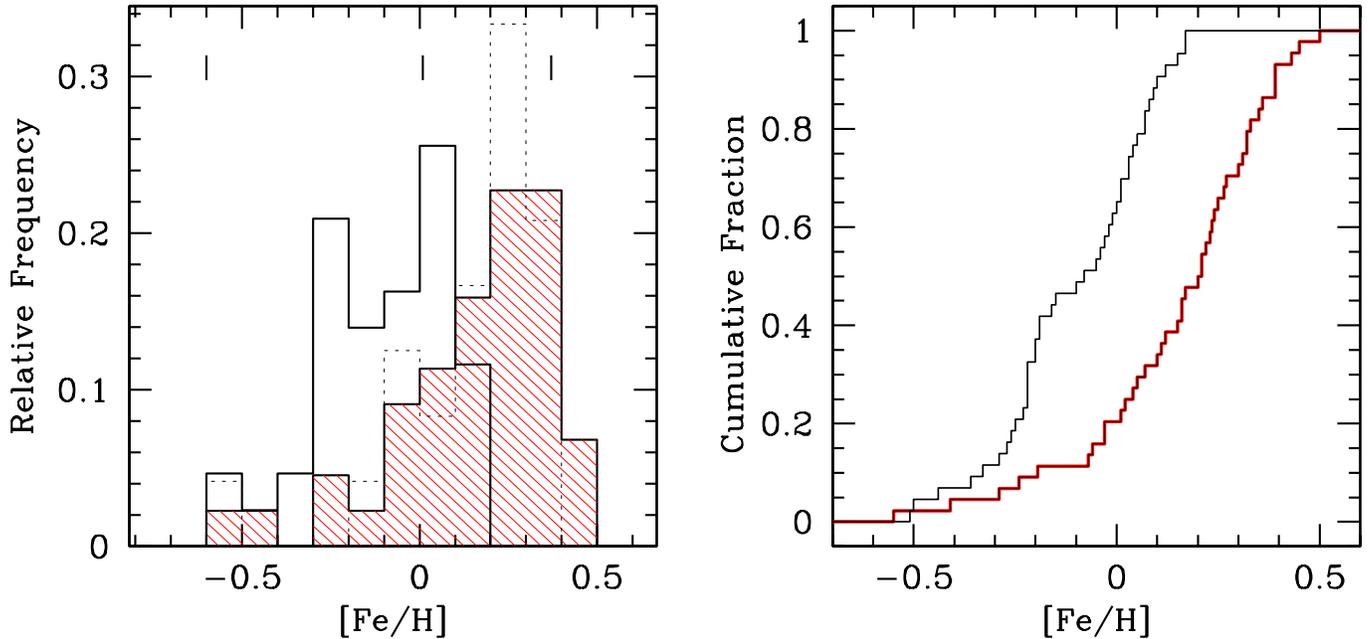}
\caption[]{{\it Left}: distribution of stars with planets (shaded histogram) compared 
with the distribution of field dwarfs presented in this paper (empty histogram). The dashed histogram represents 
the planet host sample if we consider only stars forming part of the CORALIE survey (see text for details). 
The vertical lines represent stars with brown dwarf candidate companions having minimum masses between 10 and 20\,M$_\mathrm{Jup}$. {\it Right}: The cumulative 
functions of both samples. A Kolmogorov--Smirnov test shows 
the probability of the stars' being part of the same sample is around 10$^{-7}$.}
\label{fig1}
\end{figure*}

\subsection{Observations and data reduction}

For the ``comparison sample'', spectra of S/N between 150 and 350 were 
obtained using the CORALIE high-resolution spectrograph 
($R\equiv\Delta\lambda/\lambda\sim$ 50\,000), at the Swiss 1.2-m Euler Swiss Telescope 
(La Silla, Chile) in 2000 mid November. The spectra were reduced using 
IRAF\footnote{IRAF is distributed by National Optical Astronomy Observatories, 
operated by the Association of Universities for Research in Astronomy, 
Inc., under contract with the National Science Foundation, U.S.A.} tools,
and the equivalent widths were measured by fitting a Gaussian function to each of the  lines using the ``k'' key within {\tt splot}.

Some planet host stars were also observed with CORALIE (adding a few objects to the 
study presented in Paper~I), some of them with no previous spectroscopic abundance 
determinations (see Table~\ref{tab2}). Observations of some stars with planets were 
also complemented with spectra of S/N $\sim$ 300 taken with the FEROS spectrograph ($R\sim$ 48\,000), 
at the ESO 1.52 m  Telescope (La Silla), during the nights of 2000  November 8 and 9. 
The UES spectrograph ($R \sim$ 55\,000) at the 4 m WHT (La Palma, Canary Islands) was 
also used to obtain a spectrum for the planet host star \object{HD\,190228}.

\subsection{Spectroscopic analysis}

The technique used has already been described extensively in Paper~I. 
Abundances and atmospheric parameters were determined using a standard local thermodynamic 
equilibrium (LTE) analysis with a revised version of the line abundance code MOOG 
(Sneden \cite{Sne73}), and a grid of Kurucz (\cite{Kur93}) ATLAS9 atmospheres.
As in Paper~I, we used a set of iron lines taken 
from the list of Gonzalez \& Laws (\cite{Gon00}). Here, we added one more \ion{Fe}{ii} line 
(the 5991\AA\ presented in Gonzalez et al. \cite{Gon01}) and we excluded the previously 
used \ion{Fe}{ii} line at 6432\AA\ line, because it was giving systematically low values for most 
stars.

The atmospheric parameters were obtained from the \ion{Fe}{i} and \ion{Fe}{ii} lines
by iterating until the correlation coefficients between $\log{\epsilon}$(\ion{Fe}{i}) and $\chi_l$, and 
between $\log{\epsilon}$(\ion{Fe}{i}) and  $\log{({W}_\lambda/\lambda)}$ were zero, and the
mean abundance given by \ion{Fe}{i} and \ion{Fe}{ii} lines were the same. This procedure gives 
very good results since the set of \ion{Fe}{i} lines has a very wide range of excitation potentials.
We used $\log{\epsilon_{\sun}}$(Fe)~=~7.47.

The complete results obtained for the comparison sample are presented in Table~\ref{tab1}. 
The errors in $T_\mathrm{eff}$, $\log{g}$, $\xi_t$ and [Fe/H] for a typical measure are of 
the order of 50\,K, 0.15\,dex, 0.10\,dex, and 0.06\,dex respectively. 
The number of lines used for each star, and the dispersions for \ion{Fe}{i} and \ion{Fe}{ii} 
lines are also tabulated. The masses were determined from theoretical isochrones of Schaller 
et al. (\cite{Sch92}), Schaerer et al. (\cite{Schae92}) and Schaerer 
et al. (\cite{Sch93}), using $M_{V}$ computed from Hipparcos 
parallaxes and   $T_{\mathrm{eff}}$ obtained from spectroscopy\footnote{The errors in mass represent 
simply formal errors, computed using the uncertainties in $T_\mathrm{eff}$ and M$_{V}$.}. 
Only for \object{HD\,222237} we do not present a mass, given the high error in its determination. 

In Table~\ref{tab2} we present our results of the spectroscopic analysis for stars with 
low mass companions for which we have obtained spectra. Also 
included are  the objects 
already presented in Paper~I and Santos et al. (\cite{San01}), since the values were revised; the 
differences are always perfectly within the uncertainties (usually lower than 0.02\,dex for [Fe/H]).
Errors in the parameters were computed as described in Paper~I.

For a few stars in our sample we had both CORALIE and FEROS spectra. No important systematics 
are evident in the results from the two spectrographs (due to errors in flatfields or 
background  subtraction), and we simply used a mean value of the results.

We compared our [Fe/H] determinations for the eight stars presented 
in this paper in common with the studies of Gonzalez et al. (\cite{Gon99}), 
Gonzalez \& Laws (\cite{Gon00}), and Gonzalez et al. (\cite{Gon01}) to look for any 
possible systematics. The stars and [Fe/H] obtained by these authors are 
\object{HD\,1237} (0.16), \object{HD\,16141} (0.15), \object{HD\,17051} (0.19), 
\object{HD\,38529} (0.37), \object{HD\,52265} (0.27), \object{HD\,75289} (0.28), 
\object{HD\,210277} (0.24) and \object{HD\,217107} (0.36).
The comparison shows that the formal mean difference in [Fe/H] obtained by these authors 
and from the present work is virtually 0.00\,dex; the rms of the differences is $\sim$0.01\,dex. 
The atmospheric parameters for these stars 
are also usually the same within the errors. We thus conclude that we can include their 
determinations for stars with planets in our analysis (as expected since our studies were 
similar in all aspects to theirs) without introducing any systematics, thus 
increasing  the number 
of available planetary host candidates used in the current study. The list of stars 
with planets whose determinations were made by other authors is presented in 
Table~\ref{tab3}.

\begin{figure}[t]
\psfig{width=\hsize,file=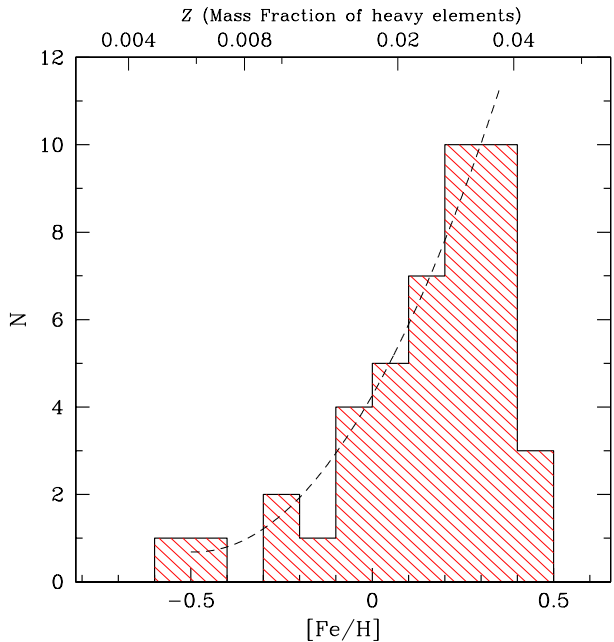}
\caption[]{The shape of the distribution of metallicities for stars with planets increases
with [Fe/H]. This represents also an increase with $Z$, the mass fraction of heavy elements $Z$. }
\label{fig2}
\end{figure}

\section{The metallicity distributions}

In Fig.~\ref{fig1} we present the [Fe/H] distributions of both samples described above. For 
the stars with planets we included both the objects presented in Tables~\ref{tab2} and \ref{tab3}.
As  is clear from the plot, stars with planets are significantly more metal rich than field 
stars without giant planet companions. While the stars-with-planets sample has a mean metallicity of 
$+$0.15 $\pm$ 0.23 the mean [Fe/H] for the field star sample is $-$0.10
$\pm$ 0.18 (here the errors represent the rms around the mean [Fe/H]). 
A Kolmogorov--Smirnov test (Fig.~\ref{fig1}, right) shows that the probability that both samples belong 
to the same population is about $\sim$10$^{-7}$. 

At this point it is important to discuss possible sources of bias.
Is the star-with-planet sample completely unbiased? If we make the same plot using only 
the stars forming part of the CORALIE sample\footnote{These include   \object{HD\,1237}, 
\object{HD\,6434}, \object{HD\,13445}, \object{HD\,16141}, \object{HD\,17051}, \object{HD\,19994}, 
\object{HD\,22049}, \object{HD\,28185}, \object{HD\,52265}, \object{HD\,75289}, 
\object{HD\,82943}, \object{HD\,83443}, \object{HD\,92788}, \object{HD\,108147}, 
\object{HD\,121504}, \object{HD\,130322}, \object{HD\,134987}, \object{HD\,168443}, \object{HD\,168746}, \object{HD\,169830}, \object{HD\,192263}, \object{HD\,210277}, \object{HD\,217107} and 
\object{HD\,222582}.} (from which the comparison stars were taken), 
the result is exactly the same (dotted histogram in Fig.~\ref{fig1}): the same general shape 
and difference is found. In fact, this result cannot be related to a selection bias since, as 
discussed above, the most important planet search programmes make use of 
volume-limited samples of stars. The only exception is \object{{\footnotesize BD}$-$10\,3166} 
(Butler et al. \cite{But00}), chosen for its high metallicity. This star was not included in our analysis. 
It is worthy of note that the six planet-host stars that were included in our
volume-limited sample (\object{HD\,1237}, \object{HD\,13445}, \object{HD\,17051}, \object{HD\,22049}, \object{HD\,28185} and \object{HD\,217107}) have a mean [Fe/H] of $+$0.12.

No important systematics are expected concerning the magnitudes of the objects. 
On the one hand, for a given colour higher [Fe/H] stars are more luminous. Also, higher 
metallicity implies more and deeper lines, and thus a more precise determination of the velocity. 
But a star with more metals is also, for a given mass, cooler and fainter. For example, doubling the metallicity 
of the Sun (i.e. increasing [Fe/H] to $\sim$ 0.30) would make its temperature decrease by 
more than 150\,K, and its luminosity by a factor of $\sim$1.2 (Schaller 
et al. \cite{Sch92}; Schaerer et al. \cite{Sch93}). As we shall see below, 
for the mass intervals for which we have a good representation of both samples, stars 
with planets are always significantly more metal-rich. Furthermore, at least in the CORALIE survey,
exposure times are computed in order to have a photon-noise error at least as low as the
instrumental errors. 

\begin{figure*}[t]
\psfig{width=\hsize,file=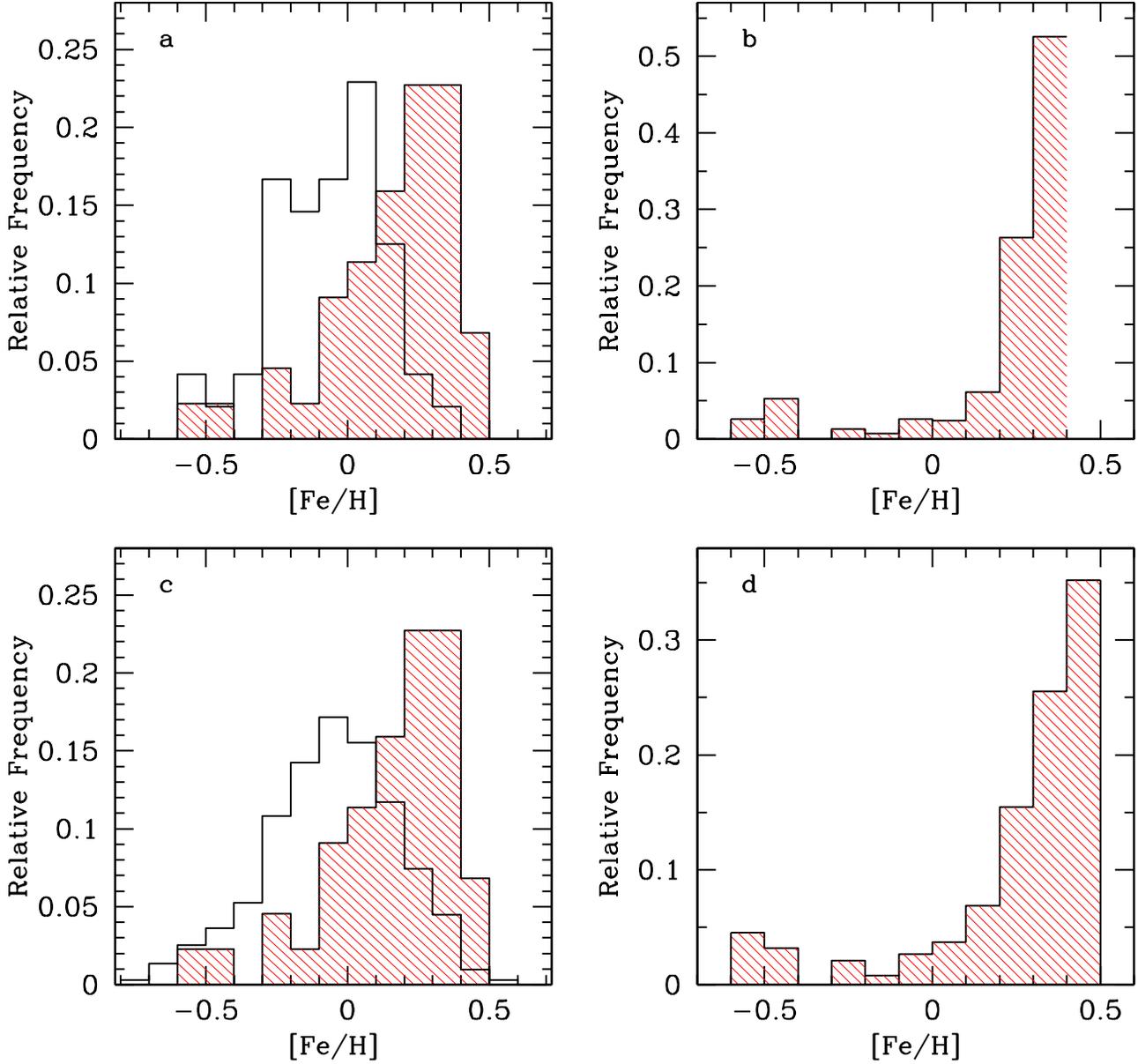}
\caption[]{Panel (a): Metallicity distribution of stars with planets (dashed histogram) compared 
with the distribution of a volume limited sample of field dwarfs (empty histogram) -- see text for 
more details; panel (b): correcting the distribution of stars with planets from the same distribution for stars in the volume-limited sample results in a even more steep rise of the planet host star distribution as a function of [Fe/H]. Note that the last bin in the planet distribution has no 
counterpart in the field star distribution; panel (c): same as (a) but for a field distribution computed using
a calibration of the CORALIE cross-correlation dip to obtain the metallicity for $\sim$1000 stars;
panel (d): same as (b) when correcting from the field distribution presented in (c). 
}
\label{fig3}
\end{figure*}

\begin{figure*}[t]
\psfig{width=\hsize,file=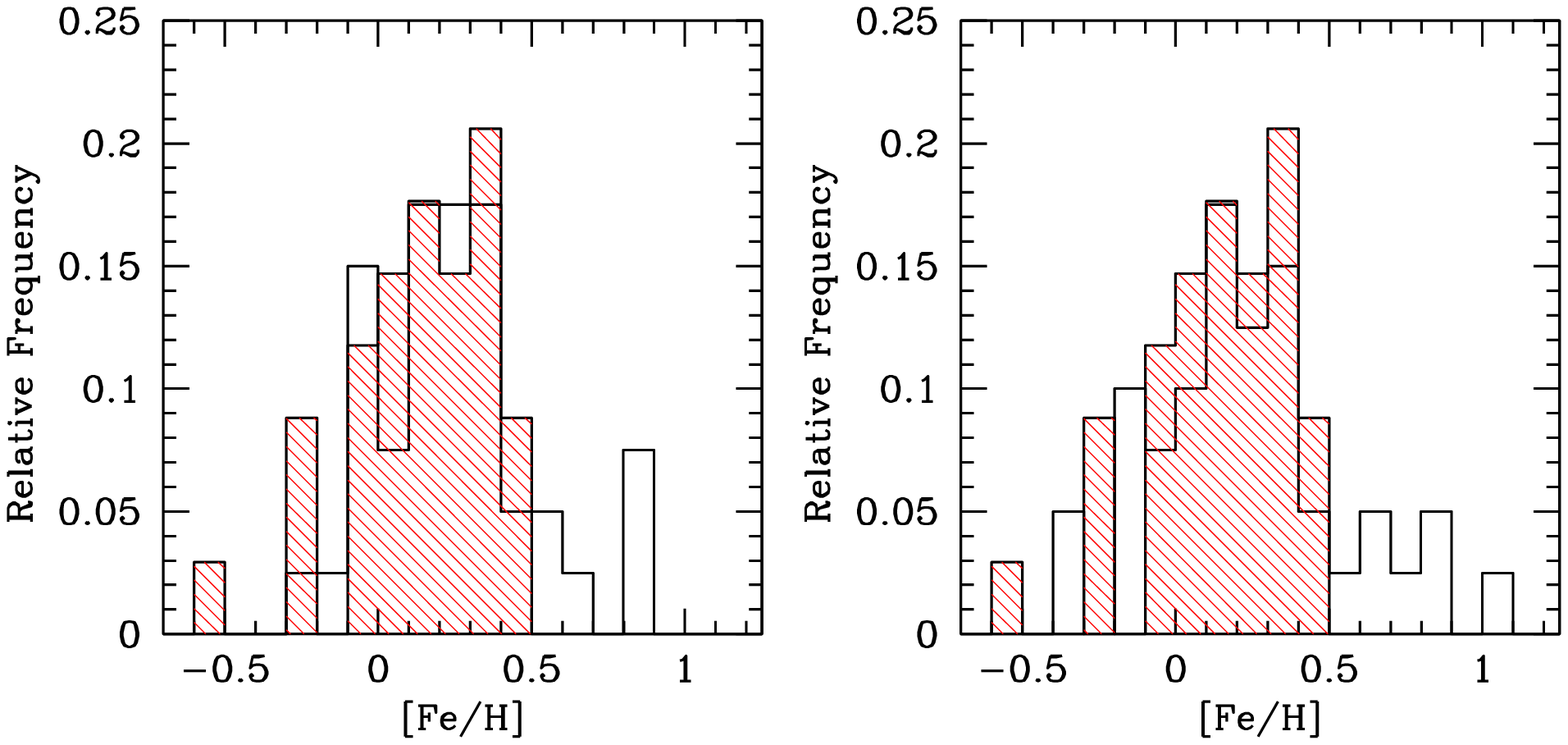}
\caption[]{Distribution for the field star sample (empty histogram) if we add 15 earth masses of iron (left) or a quantity of iron proportional to $Z$ (right) -- in the latter case, 0.003$\times$$Z$. The distributions peak at about the same value as the distribution of stars with planets (dashed histogram), but have completely distinct forms and extents in [Fe/H]. See the text for more details.}
\label{fig4}
\end{figure*}

Given the uniformity of the study, we may conclude that the plot in Fig.~\ref{fig1} represents a proof 
that the stars now observed  with giant planets are, on average, more metal-rich than field stars. 
For the record, it is also interesting to verify that the mean value of [Fe/H] obtained by Favata et al. (\cite{Fav97}) for their volume-limited sample of G-dwarfs is $-$0.12, which means that there are
almost no systematics between the method used by these authors and that used for the current study.

Also remarkable in Fig.~\ref{fig1} is the interesting position of the three
 low-mass (10\,M$_\mathrm{Jup}$$<$M\,$\sin{i}$$<$20\,M$_\mathrm{Jup}$) brown-dwarf candidates.
Although it is too early to arrive at any firm conclusions, the position of one of them (\object{HD202206}),
with [Fe/H] = $+$0.37 is strongly suggestive of
 a common origin with the lower-mass planets. On the opposite
side of the distribution there is \object{HD\,114762} ([Fe/H] = $-$0.60), the most metal-poor object among 
all those studied in this paper. This ``dispersion'' might be interpreted as a sign that the frontier 
between brown dwarfs and massive giant planets is very tenuous (and probably overlaps)  with regard to the mass
limit; we are possibly looking at results of different formation processes (e.g. Boss \cite{Bos00}).

\section{How to explain it?}

The problem concerning the significance of the high-metallicity trend observed for stars with 
planets being ``solved'', our enquiry
 is turned to the cause of these differences and to 
the implications this result might have for planetary system formation and evolution 
scenarios.

\subsection{Comparing the distributions}

One very interesting feature of the distribution of stars with planets is well illustrated in 
Fig.~\ref{fig2}. The shape of this distribution is well described by a quadratic rising 
with [Fe/H] (up to [Fe/H] $\sim$ $+$0.35 -- dashed curve in the diagram). In other words, the 
distribution is rising with $Z$ (the mass fraction of heavy elements) up to a value of 0.04,
falling then abruptly. This sharp cut-off in the distribution of stars with planets is probably due 
to the fact that we may be looking at a limit on the metallicity of the stars in the solar 
neighbourhood.

Clearly, the rise is expected if we consider that the probability of forming planets
is proportional to the metallicity content. In one sense, an increase in the number of dust 
particles ($N$) in the young disc will increase the probability of a shock by $N^2$. 
The rate of formation of planetesimals, and thus the probability of creating a planet core with 
sufficient mass to accrete gas from the disc before it disappears, will increase quadratically 
with $Z$ (considering that there is a linear relation between these two variables).


But this model is probably too simple. Other, more complicated mechanisms intervene, however, 
such as the fact that an increase in [Fe/H] will also probably increase the 
opacity of the disc, changing its physical conditions (e.g. Schmidt et al. 
\cite{Sch97}), such as the accretion rate or its vertical structure. The current 
result may be used in exactly this way in constraining
these mechanisms.

On the other hand, the distribution of [Fe/H] in a volume-limited sample of stars
(with and without planet hosts) apparently decreases with increasing [Fe/H] for
[Fe/H] $>$ 0.0 (e.g. Favata et al. \cite{Fav97}). Taking this effect into account would 
result in an even steeper rise of the relative frequency of stars with planets with the metallicity.

This is exactly what we can see from Fig. \ref{fig3}. In panels (a) and (c)
we present a comparison between the [Fe/H] distribution of stars with planets (shaded bars) and the same distribution for two volume-limited samples of dwarfs
(empty bars). The volume-limited distribution in panel (a) was taken
from this paper (objects presented in Table\,\ref{tab1} plus the six 
stars with planets making part of our volume limited sample -- see Sect. 2.1 for details).
For panel (c), the distribution represents the metallicities of about 1000 stars in the CORALIE planet-search sample as determined from a calibration of the CORALIE cross-correlation function surface 
(Santos et al., in preparation)\footnote{The cross-correlation dip can be seen as an average spectral 
line; as shown by Mayor (\cite{May80}) and Pont (\cite{Pon97}) its surface is well correlated with [Fe/H] and $(B-V)$. A calibration of its surface with these two variables 
can in fact be used to obtain in a very simple way precise metallicity 
estimates, at least in a statistical sense. In this particular case, 
the calibration used to construct this ``comparison'' sample uses as ``calibrators'' the 
stars presented in Tables\,\ref{tab1} and \ref{tab2}, and thus the results are in the same 
[Fe/H] scale as our spectroscopic values. The precision of the calibration is remarkable (rms$\sim$0.01\,dex). The resulting distribution is quite symmetric, and peeks at 
$\sim$$-$0.05 dex.}. Since this two field star distributions represent the
actual distribution of metallicities in the solar neighborhood (within statistical errors), we can use them to estimate the ``real'' percentage of stars with planets for each metallicity bin that we would find if the metallicity distribution of field stars was ``flat'' -- panels (b) and (d). 
As we can see from these plots, there is a steep rise of the percentage of stars with planets with [Fe/H] (we will concentrate on panel (d) given that it presents better statistics than panel (b)). Actually, the plot shows that more than 75\% of the stars with planets would have [Fe/H]$>$$+$0.2\,dex. But even more striking is the 
fact that $\sim$85\% of the surface of the distribution falls in the region of 
[Fe/H]$>$0.0. Although this is a preliminary result, the plots leave no doubts about the strong 
significance of the increase in the frequency of known planetary systems found with the metallicity 
content of their host stars. 

We note that for the low metallicity tail of the distribution, this result is based on
relatively poor statistics. The small relative number of stars with 
planets found with low [Fe/H] values is, however, perfectely consistent with recent studies 
that failed to find any planetary host companions amid the stars in the metal-poor globular 
cluster \object{47\,Tuc} (Gilliland et al. \cite{Gil00}).

\subsection{A simple ``pollution'' model}

We tried to determine whether it would be possible to obtain a metallicity distribution 
for the stars without planets similar to that of stars with 
extrasolar planets by simply 
adding material to their convective envelopes. To make this simple model, we computed the convective 
zone mass for each star using Eq. (5) and (6) of Murray et al. (\cite{Mur01}) -- 
metallicity-dependent relations for $M<$ 1.2M$_{\sun}$; only stars in this mass range 
were used, since the comparison sample is not complete for higher masses. We are supposing 
that the ``pollution'', if significant, has taken place after the star reaches the ZAMS. As discussed 
in Paper~I, if we consider  the pollution to occur in a pre-main-sequence 
phase, higher quantities of material would be needed, since the convection envelope masses are bigger 
(D'Antona \& Mazzitelli \cite{Dan94}). In fact, recent results show that gas-disc 
lifetimes can be higher than previously thought (up to 20 Myr -- Thi et al. \cite{Thi01}), 
and thus we can probably expect significant pollution when the star is already on the main sequence.

Two models were then constructed: the first was built on the supposition that the quantity of material falling into the star 
is the same for all the objects. A second, and possibly more realistic model, considered  
the pollution to be  proportional to $Z$. The quantity of ``added material'' was chosen so that the 
distributions (stars with and without planets) peak at about the same value of [Fe/H]. 

The results are shown in Fig.~\ref{fig4} (open bars), compared with the results 
obtained for stars with planets within the same mass regime. As we can see, 
the shape of the planet sample, particularly the sharp cutoff at high metallicity, is not correctly 
obtained for any of the models. 
The resulting samples are more symmetric, but most of all, in both models there are always 
a few points that end up with extremely high metallicities ($>$ 0.8\,dex). This is particularly 
unpleasant, since the inclusion of higher-mass stars (such as
 some of the observed exoplanet hosts) 
would even boost some bins to higher values, given that the mass of the convection zone is supposed to 
be lower. As discussed by Murray et al. (\cite{Mur01}), however, Li and Be observations suggest 
that the total ``mixing'' mass increases after 1.2 M$_{\sun}$, eventually reducing this problem. 
Note also that the strong cut-off in [Fe/H] for the planet host star sample remains, even 
though this time we have not included stars with $M~>~$1.2 M$_{\sun}$. This shape is not at all
reproduced by our computations.

We caution, however, that we are using very simple models to make this calculation. 
The inclusion of more parameters, such as
 a time- or disc-mass-dependent accretion (which might be 
related to the stellar mass), or a mass-dependent size of the central disc hole (e.g. Udry 
et al. \cite{Udr01}), might modify the results. In any case, we have shown that a simple ``pollution'' model cannot explain the observations.

\begin{figure}[t]
\psfig{width=\hsize,file=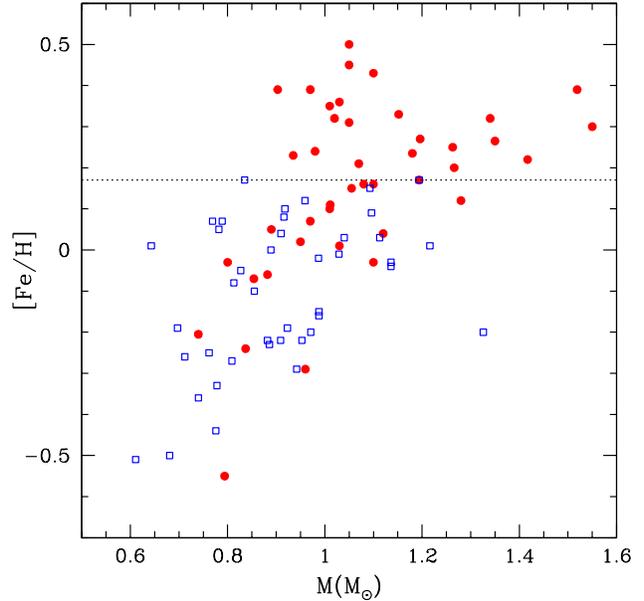}
\caption[]{A plot of [Fe/H] as a function of mass for planet host stars 
(filled circles) and ``comparison'' stars (open squares). The line represents the upper 
limit for the field star [Fe/H]. See text for more details.}
\label{fig5}
\end{figure}

\subsection{The mass dependence}

It has been suggested (Laughlin \& Adams \cite{Lau97}) that if the pollution scenario is correct 
one would be able to see a trend of [Fe/H] with stellar mass, since the 
high-mass stars have tiny 
convective envelopes, and thus a higher probability of having a noticeable increase in [Fe/H] 
(here we have chosen stellar mass and not the mass of the convective 
shell, since if the pollution scenario is correct, the original mass for the convection zone for the 
stars with planets would not be correctly calculated). In Fig.~\ref{fig5} we present such a plot. 

There are three features in the plot that deserve a comment. First, the region for $M\ge$ 1.2 M$_{\sun}$ 
has a very low number of stars without planets. This is because
 in a volume-limited sample of 
stars, there are very few dwarfs in this mass regime (the IMF decreases with mass). However, these 
stars are bright, and thus easier targets for planet searches, which
explains the high number of planetary 
candidates in this region. On the other hand, for masses $M\le$ 0.8 M$_{\sun}$, we see the opposite effect: 
stars are fainter, and thus difficult targets for planet searches, but are more numerous in a 
volume-limited sample. There are probably, however, two more biases in the diagram. In fact, the comparison 
sample was cut in $(B-V)$. For a given temperature, a higher-metallicity star
also has higher mass. Thus, 
stars in the upper part of the diagram (more metal-rich) are moved to the 
righthand side relative to 
stars in the lower part. For a fixed $T_\mathrm{eff}$, 0.2 dex in metallicity imply $\sim$ 0.05\,$M_{\sun}$. 
On the other hand, if we take a given volume-limited sample of dwarfs, 
lower-mass dwarfs tend to be older 
than high-mass dwarfs; in proportion there are thus fewer metal-rich dwarfs
in the ``high''-mass region of the 
plot. These facts might explain the lack of stars with planetary companions presenting $M\ge$ 1.2\,M$_{\sun}$ 
and low [Fe/H]. 
 
In any case, the most striking thing in the diagram is the fact that all comparison stars have 
[Fe/H] $\le$ 0.17\,dex, while about 55\% of stars with planets have metallicity above that value. 
Furthermore, contrary to former studies (Laughlin \cite{Lau00}) this diagram does not suggest 
that the metallicity of the upper envelope of stars with planets is increasing with stellar mass. 
This point renders unlikely the possibility that the high metallicity 
of these stars is due to pollution. As discussed above, however, for stellar masses higher 
than $\sim$ 1.2\,M$_{\sun}$, the total ``mixing'' mass may increase. This may explain to some 
extent the lower value of the upper envelope of [Fe/H] for masses above this limit, but it is not certain 
that it is enough to explain the difference. An extension of our comparison 
sample for stars with $M>$ 1.2\,M$_{\sun}$ would be very useful.

As mentioned, our results and interpretations are opposed to those presented by Laughlin 
(\cite{Lau00}), who used photometric indices to compute the metallicities for a sample of stars 
with planets, and for a large volume-limited sample of dwarfs used as ``comparison''. We believe 
that there are two important reasons for the difference. First, the trend he finds in [Fe/H] vs. stellar 
mass is mainly due to the few ``low''-[Fe/H] planet host stars having $M<$ 1 M$_{\sun}$. On the other hand, 
in his plot there are a few very high metallicity ``comparison'' stars in the region of $M<$ 1.1 M$_{\sun}$. 
If we believe in the results presented in this paper, the probability that these stars harbour planets is 
very high. His sample might in fact be biased by the presence of planet host stars in his ``comparison'' sample.

\begin{table}
\caption[]{Planetary and orbital parameters used in Fig.~\ref{fig6} and \ref{fig7}.
Stars with multiple companions are not included. The values were compiled from the literature.}
\begin{tabular}{llccccc}
\hline
\noalign{\smallskip}
HD      & Star  & M\,$\sin{i}$       & Period & a      & e \\
number  &       & ($M_\mathrm{Jup}$) & (days) & (A.U.) &   \\
\hline \\
1237  &\object{GJ\,3021}    &3.51 &133.7 &0.494 &0.51\\
6434  &\object{HD\,6434 }    &0.48 &22.1 &0.15 &0.295\\
10697 &\object{109\,Psc  }  &6.60 &1072 &2.12 &0.12\\
12661 &\object{HD\,12661 }  &2.93 &264.0 &0.789 &0.33\\
13445 &\object{Gl\,86    }  &3.77 &15.8 &0.11 &0.04\\
16141 &\object{HD\,16141 }  &0.23 &75.8 &0.35 &0.28\\
17051 &\object{$\iota$\,Hor}   &2.36 &311.3 &0.93 &0.22\\
19994 &\object{HD\,19994 }  &1.88 &454.2 &1.23 &0.2\\
22049 &\object{$\epsilon$\,Eri }   &0.84 &2518 &3.4 &0.6\\
28185 &\object{HD\,28185}  & 5.6 & 385 & 1.0 & 0.06 \\
37124 &\object{HD\,37124 }  &1.04 &154.8 &0.55 &0.31\\
38529 &\object{HD\,38529 }  &0.81 &14.3 &0.129 &0.27\\
46375 &\object{HD\,46375 }  &0.26 &3.02 &0.041 &0.02\\
52265 &\object{HD\,52265 }  &1.08 &119.2 &0.5 &0.35\\
75289 &\object{HD\,75289 }  &0.44 &3.48 &0.04 &0.065\\
75732A &\object{55\,Cnc  }   &0.82 &14.66 &0.12 &0.03\\
80606 &\object{HD\,80606}  & 3.51 & 111.8 & 0.44 & 0.93\\
89744 &\object{HD\,89744 }  &7.54 &265.0 &0.91 &0.7\\
92788 &\object{HD\,92788 }  &3.77 &340.8 &0.97 &0.36\\
95128 &\object{47\,UMa   }  &2.57 &1084 &2.1 &0.13\\
108147 &\object{HD\,108147}  &0.35 &11.05 &0.10 &0.57\\
117176 &\object{70\,Vir   }  &7.75 &116.7 &0.48 &0.4\\
120136 &\object{$\tau$\,Boo} &4.29 &3.31 &0.047 &0.02\\
121504 &\object{HD\,121504 } &0.89 &64.62 &0.32 &0.13\\
130322 &\object{HD\,130322 } &1.04 &10.72 &0.088 &0.044\\
134987 &\object{23\,Lib    } &1.57 &259.6 &0.81 &0.24\\
143761 &\object{$\rho$\,CrB} &1.15 &39.6 &0.23 &0.07\\
145675 &\object{14\,Her    } &5.65 &1650 &2.84 &0.37\\
168746 &\object{HD\,168746 } &0.25 &6.41 &0.066 &0.00\\
169830 &\object{HD\,169830 } &3.04 &229.9 &0.82 &0.35\\
177830 &\object{HD\,177830 } &1.26 &391.6 &1.1 &0.41\\
186427 &\object{16\,Cyg\,B } &1.75 &804.4 &1.61 &0.67\\
187123 &\object{HD\,187123 } &0.60 &3.10 &0.042 &0.01\\
190228 &\object{HD\,190228 } &5.03 &1161 &2.3 &0.5\\
192263 &\object{HD\,192263 } &0.75 &24.13 &0.15 &0.00\\
209458 &\object{HD\,209458 } &0.68 &3.52 &0.047 &0.00\\
210277 &\object{HD\,210277 } &1.31 &435.6 &1.09 &0.34\\
217014 &\object{51\,Peg    } &0.47 &4.23 &0.05 &0.00\\
217107 &\object{HD\,217107 } &1.34 &7.11 &0.071 &0.14\\
222582 &\object{HD\,222582 } &5.44 &575.9 &1.35 &0.71\\
\noalign{\smallskip}
\hline
\end{tabular}
\label{tab4}
\end{table}

If we concentrate on the mass interval for which we have both stars with planets and comparison 
stars (region with 1.2\,M$_{\sun} > M > 0.7$\,M$_{\sun}$), we obtain a mean [Fe/H] = $+$0.13 for the 
star with planet sample, and $-$0.08 for the comparison star sample. This makes a difference of about 
0.21\,dex, for a mean stellar mass of 0.95\,M$_\odot$.
To pollute the convective envelope of a 
0.95 solar mass star containing about 0.04 M$_{\sun}$ in the convection zone (D'Antona \& Mazzitelli 1994) 
in order to obtain the observed difference, one would need about 10 earth masses of pure 
iron. Considering the composition of C1 chondrites, this would mean $\sim$ 5 times more in 
silicate material, a value that seems excessively high. The addition of giant planets, 
also very rich in gas, would need even more material. For example, the addition of Jupiter to the 
present-day Sun would only be able to increase its metallicity by $\sim$ 0.05\,dex, while adding two jupiters 
would increase 0.08\,dex (e.g. Gonzalez \cite{Gon98a}). We are supposing here that Jupiter has a
rocky core, which is not necessarily true (Guillot \cite{Gui99}).

\begin{figure*}[t]
\psfig{width=\hsize,file=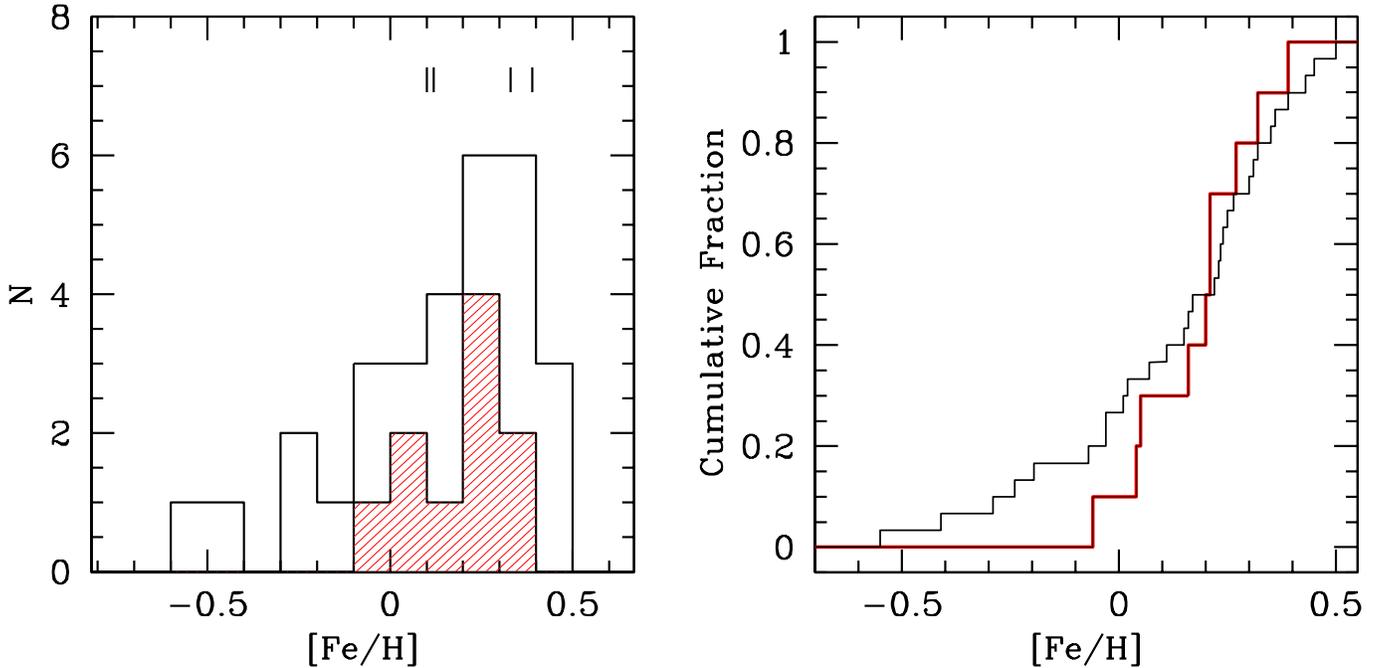}
\caption[]{{\it Left}: distribution of [Fe/H] for stars with planets orbiting with semi-major axes 
lower than 0.1~AU (shaded histogram), and with 
semi-major axes greater than 0.1~AU (open histogram). The vertical lines denote stars 
with more than one planetary-mass companion. {\it Right}: cumulative fraction of the [Fe/H] distribution for
the long (thin line) and short (thick line) period systems. Althought some trend is suggested
regarding a possible higher metallicity for the stars with short period planets, a Kolmogorov--Smirnov 
test gives a probability of $\sim$0.7, which is not significant.}
\label{fig6}
\end{figure*}

Murray et al. (\cite{Mur01}) have reported evidence that all stars in the solar 
neighbourhood have accreted about 0.4 Earth masses of  iron, suggesting that ``terrestrial-type 
material is common around solar type stars.'' This seems to be particularly evident in the objects with
$M>$ 1.3--1.4\,M$_{\sun}$, which is in accordance  with  expectations (Laughlin et al. \cite{Lau97}). 
If confirmed, and as discussed by these authors, this result further stresses the difficulty of 
obtaining the [Fe/H] differences we observe considering a pollution model: 0.4 Earth masses of 
iron are definitely not able to explain the observed differences between stars with planets and 
stars ``without'' planets.

\subsection{Using evolved stars as a test}

One other interesting note can be added by looking at the ``star-with-planet'' sample. 
If the pollution scenario were correct, we would expect evolved stars (sub-giants) having 
planetary companions to have lower metallicities than their dwarf counterparts, since the 
convection envelope increases 
in mass as the star evolves off the main sequence (Sackmann et al. \cite{Sac93}), diluting its metallicity content. 

There are eight stars in Tables~\ref{tab2} and \ref{tab3} having spectroscopic $\log{g}$ values below 
$\sim$ 4.10 dex (probably already slightly evolving away from
 the main sequence): \object{HD\,10697} 
(3.96), \object{HD\, 38529} (4.01), \object{HD\,117176} (3.90), \object{HD\,143761} (4.10), 
\object{HD\,168443} (4.10), \object{HD\,169830} (4.04), \object{HD\,177830} (3.32) and 
\object{HD\,190228} (4.02). If we compute the mean [Fe/H] we obtain a value of $+$0.08 $\pm$ 0.25 
(with values going from $-$0.29 to $+$0.39), only slightly lower than the value of $+$0.15 $\pm$ 0.22 for 
the sample comprising only  stars with $\log g > 4.10$ (values going from $-$0.55 to $+$0.50). 
As mentioned also by Gonzalez et al. (\cite{Gon01}), this is even more striking when we 
notice that the higher-metallicity star in this ``low $\log{g}$'' sample (\object{HD\,177830}) is 
that having the lower surface gravity.

\section{Metallicity and orbital parameters}

Given that we already have more than 40 extrasolar planets with
high-precision metallicity determinations, we can start to think about
looking for possible trends in [Fe/H] with planetary mass, semi-major axis 
or period, and eccentricity. 

Gonzalez (\cite{Gon98a}) and Queloz et al. (\cite{Que00}) have suggested that 
stars with short-period planets (i.e. small semi-major axes) may be particularly metal-rich,
even amongst the planetary hosts.
The number of planets that were known by that time was, however, not enough
to arrive at a definitive conclusion. Here we revisit that discussion with much 
better statistics.

In Fig.~\ref{fig6} we present a plot of the distribution of planets for two
orbital ``radius'' regimes: stars with planets orbiting with semi-major axes lower than 
0.1~AU (shaded histogram), and with semi-major axes greater 
than 0.1~A.U (open histogram). The values were taken from Table~\ref{tab4}. 
Stars with multiple companions are denoted by the vertical lines and were not included in 
the histograms. These include, among the stars for which we have [Fe/H] values available, the multiple systems around \object{$\upsilon$\,And} (Butler et al. \cite{But99}), \object{HD\,83443} (Mayor et al. \cite{May01}), 
\object{HD\,168443} (Udry et al. \cite{Udr01}), and \object{HD\,82943} (Mayor et al. \cite{May01b}).

\begin{figure*}[t]
\psfig{width=\hsize,file=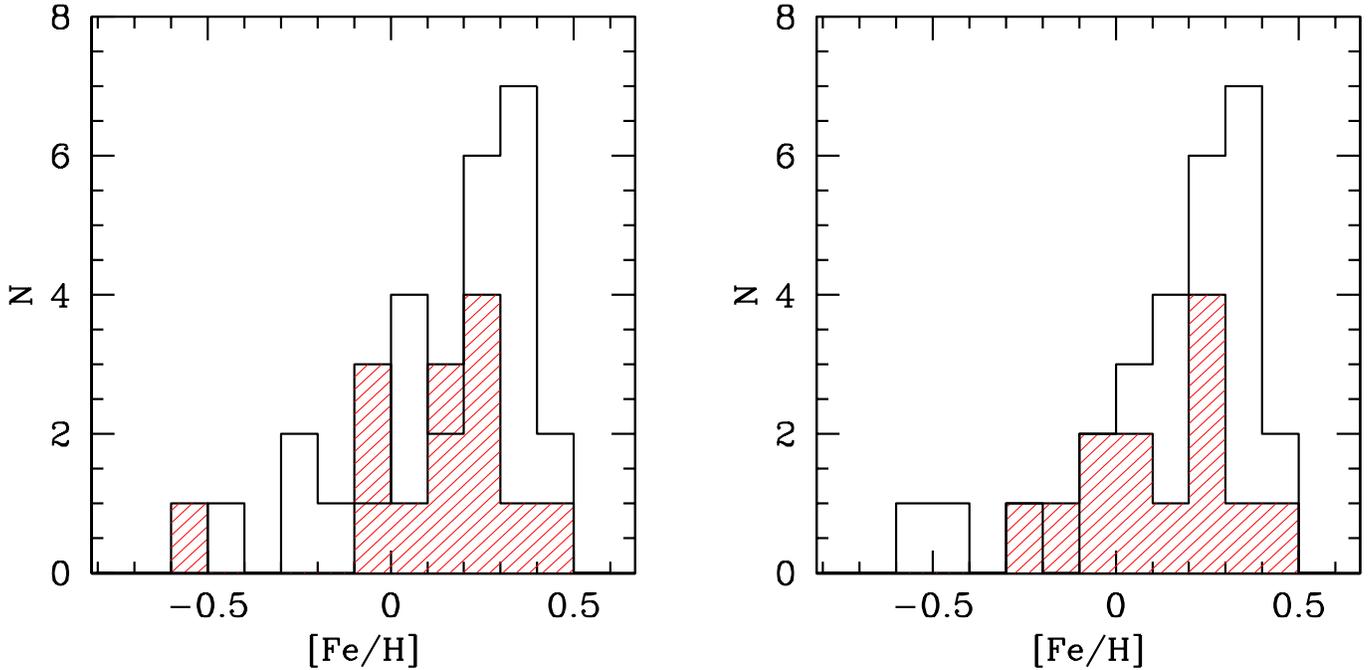}
\caption[]{{\it Left}: distribution of [Fe/H] for stars with planets having masses lower 
than 1\,$M_\mathrm{Jup}$ (shaded histogram), and greater than 1\,$M_\mathrm{Jup}$ (open histogram). 
{\it Right}: Similar to the left panel, but for eccentricity (limit at 0.1). In both cases, 
a Kolmogorov-Smirnov test results in a non-significative probability of being two different populations.}
\label{fig7}
\end{figure*}

Some trend can be seen in the plot (right panel) suggesting that stars with short period planets 
might be more metal-rich then their long-period counterparts.
A Kolmogorov--Smirnov test, however, gives a probability that both 
distributions make up part of the same sample of $\sim$75\%. We have tried to make the plot using 
different limits of the semi-major axes; no further conclusions can be drawn.
It is interesting to note that one of the longest-period planets discovered up to now (around 
the star \object{14 Her}) has a value of [Fe/H] = $+$0.50. 

Similar conclusions are drawn respecting the eccentricity or the companion mass (Fig.~\ref{fig7}).
The current results do not suggest any clear difference in [Fe/H] between high- and low-mass/eccentricity 
planets. If there is some trend, then more data (and probably even more precise abundances) are needed 
to permit conclusions. In the same way, no conclusions are possible regarding the position of the planetary systems in the plot (Fig.~\ref{fig6}).

\section{Concluding remarks}

We have presented for the first time a uniform and unbiased comparison between the metallicity 
distributions of a sample of stars with planets and a sample of stars ``without'' planets. 
The main results can be summarized as follows:

\begin{itemize}

\item The currently known stars with planets are substantially
 metal-rich when compared with non-planetary 
host dwarfs. The mean difference in [Fe/H] is $\sim$0.25\,dex and is clearly significant.

\item The shape of the metallicity distribution of stars with planets has a very clear rise with 
[Fe/H], i.e. with $Z$, the mass fraction of heavy elements. This result, although still
based upon ``only'' 44 objects, may help put constraints on the models and conditions leading to 
giant-planet formation.  

\item The clear cut-off seen for the star-with-planet distribution at [Fe/H] $\ge$ $+$0.5
may be interpreted as a ``limit'' for the metallicity of solar-type dwarf stars in the solar 
neighbourhood; it is very difficult to explain it in terms of a convective envelope ``pollution'' 
scenario.

\item The metallicity of stars with planets cannot be explained by a simple ``pollution'' model.
The differences in [Fe/H] between stars with planets and field stars does not seem
to be explained by such mechanisms, but needs an original (cloud) metallicity ``excess''. 
Otherwise huge quantities of iron-rich material would have to be engulfed, which is not easy to
explain in face of the current observations. Our results do not rule
out, however, planet accretion as an ongoing process in the early stages of
planetary formation (during which the convective envelopes are more massive).

\item An analysis of the planetary orbital parameters does not reveal any clear trends with [Fe/H].
This point might give further support to the ``primordial'' origin of the high metallicity of
stars with planets (e.g. Gonzalez \cite{Gon98a}; Del Popolo et al. \cite{Pop01}). On the other hand, 
we cannot exclude the possibility that some trend may appear in the future, e.g. when some real solar 
system analogues are found.

\end{itemize}

The current result concerning the high metallicity of stars with planets may represent the 
simple fact that the higher the metallicity of the star, the higher will be the probability 
that planet formation will occur. 

Recently, Israelian et al. (\cite{Isr01}) have discovered that the planet host star \object{HD\,82943}
has $^6$Li in its atmosphere. This discovery is interpreted as evidence of the infall of a planet into
the star, most probably by planet disruption (Rasio \& Ford \cite{Ras96}). As discussed by the authors, 
however, the high [Fe/H] value of this star is not necessarily related to the infall of a planet, as 
this would not be able to enhance the [Fe/H] value by more than a few hundredths of a dex. This is 
exactly what was found for the pair \object{16\,Cyg} A and B (Laws \& Gonzalez (\cite{Law01}). 

In fact, the present results do not exclude the possibility that pollution may play a role 
(eventually important in some cases, 
such as for the most massive dwarfs), but rather that it is not the key process leading to 
the observed high metallicity of the planet host stars. Recent work by Murray et al. (\cite{Mur01})
supports the idea that pollution, on a small scale, is very common among
the  stars of the solar neighbourhood. 
If confirmed, however, this result does not change in any way
 the conclusions of this paper, since their models predict ``pollution'' of the order of 0.4 
 Earth masses of iron, only noticeable in stars with $M>$1.5\,M$_{\sun}$.

One remarkable point that becames evident from the current work is the fact that our Sun occupies 
a ``modest'' position in the low [Fe/H] tail of the metallicity distribution of stars with planets. 
A look at the orbital parameters in Table~\ref{tab4}, however, tells us that, to date, no ``real'' 
Solar System analogs were found. This lead us to speculate about possible different formation histories 
for these systems. Note, however, that we cannot draw any conclusions until other Solar System analogs 
are found.

The future addition of more stars in both samples studied here, and the detailed analysis of other 
chemical elements (in particular the light elements Li, Be and B) are of particular interest 
for the future, both to consolidate  current results, and  to  understand better 
the multitude of different planetary systems known today. The direct comparison of other elemental abundances (e.g. $\alpha$-elements, C, or O) between stars with planets and ``single'' field dwarfs 
will, on the other hand, not be an easy task given the
apparent lack of stars with high [Fe/H] having no giant planet companions. 
Meanwhile, we expect that theoretical models of disc evolution and (giant) planetary formation
will be constructed to explain in a detailed way the physics beyond the current results.

\begin{acknowledgements}
  We wish to thank the Swiss National Science Foundation (FNSRS) for
  the continuous support for this project. 
  We would also like to thank Nami Mowlavi for help computing
  masses for the stars, and to Terry Mahoney for the useful english comments.
  Support from Funda\c{c}\~ao para a Ci\^encia e Tecnologia, Portugal,
  to N.C.S. in the form of a scholarship is gratefully acknowledged.
\end{acknowledgements}


\end{document}